# Robust Learning Equilibrium

Itai Ashlagi            Dov Monderer            Moshe Tennenholtz

Faculty of Industrial Engineering and Management
Technion–Israel Institute of Technology
Haifa 32000, Israel

## Abstract

We introduce *robust learning equilibrium* and apply it to the context of auctions.

## 1 INTRODUCTION

Learning in the context of multi-agent interactions has attracted the attention of researchers in psychology, economics, artificial intelligence, and related fields for quite some time (Kaelbling, Littman, & Moore (1996); Erev & Roth (1998); Fudenberg & Levine (1998)). Much of this work uses repeated games (e.g. Claus & Boutilier (1997); Kalai & Lehrer (1993); Hart & Mas-Colell (2001); Conitzer & Sandholm (2003)) and stochastic games (e.g. Littman (1994); Hu & Wellman (1998); Brafman & Tennenholtz (2002); Bowling & Veloso (2001); Greenwald, Hall, & Serrano (2002)) as models of such interactions. Roughly speaking, work on learning in games consists of two different paradigms. One paradigm, which extends upon Bayesian learning (see Kalai & Lehrer (1993)) deals with situations where there exists a known prior distribution on the set of possible games. The other paradigm, which is more consistent with work on reinforcement learning in computer science, cognitive psychology, adaptive control, and related disciplines, deals with non-Bayesian models, where an agent aims to adapt to its environment while not having exact prior distribution on the environment's structure. Indeed, most recent work on multi-agent learning in computer science fits this (latter) paradigm, which is the paradigm that will be adopted in the rest of this paper.

Consider a class of one-stage games $\Gamma_\omega$. One of the games in this class is chosen and is played repeatedly. Had the agents known the chosen game they would have used algorithms that are in equilibrium in the associated repeated game. However, the agents do not have information about the chosen game, and can only partially observe the other agents' actions. Most of the work on learning in games is concerned with the understanding of learning procedures that **if** adopted by the different agents will converge at end to an equilibrium of the (one-stage) chosen game. The learning algorithms themselves are not required to satisfy any rationality requirement; it is what they converge to, **if** adopted by **all** agents that should be in equilibrium. In most of these algorithms, a unilateral deviation by a strategic agent from its learning algorithm may make him better off in some of the possible games in the class. That is, the learning algorithms **themselves** are not in equilibrium. In contrast, in a *Learning Equilibrium* (Brafman & Tennenholtz (2004)) a unilateral deviation is not beneficial to the deviator at every possible game in the class. That is, in a learning equilibrium the learning algorithms are repeated-game equilibrium strategies for every game in the class.

The learning equilibrium perspective is in fact an extension of the classical single agent reinforcement learning paradigm to the context of multi-agent systems. In a classical reinforcement learning setup, such as Q-learning (Watkins (1989)), we are given a class of models and our aim is to devise an algorithm that would be as successful as the optimal algorithm one could have devised had he known the model to start with. In fact, the "signature" of learning (when compared to optimization) is that we aim for success for *any* model taken from a given set of possible models, although initially we do not know what the actual model is. When considering the multi-agent setup the set of possible models is the set of possible games, and instead of requiring optimality at every possible model we require that the learning algorithms will be in equilibrium at every possible game.

This paper consists of three logical parts.

The first part extends the domain of definition of learning equilibrium to allow general monitoring structure and initial private information of the agents.

The second part introduces robust learning equilib-rium. We define a robust learning equilibrium as a learning equilibrium which is immune to failure of the agents to follow their algorithms for some finite time. The relation between learning equilibrium and robust learning equilibrium resembles the relation between equilibrium and sub-game perfect or sequential equilibrium. However, we consider also the issue of system failures in providing the correct information to the agents. We model that by explicitly considering failure patterns as part of our description. In this extended setting learning algorithms which form a robust learning equilibrium recover from any of the possible failure patterns. That is, in this context, a robust learning equilibrium is immune against both agents' failures and system failures.

In the third part we initiate the study of robust learning equilibrium in auctions. We present a family of natural learning algorithms and prove that when all agents adopt any of the algorithms in that family, a robust learning equilibrium is formed. Moreover, this robust learning equilibrium is immune against arbitrary failure patterns. Our results are obtained under an extremely weak monitoring structure where an agent can observe only winning bids. Our study also complements work on learning in auctions. While previous work (Hon-Snir, Monderer, & Sela (1998)) has shown learning algorithms that converge to equilibrium, our work is the first to show learning algorithms that are in equilibrium. Moreover, these learning algorithms are robust and immune to system failures.

## 2 ROBUST LEARNING EQUILIBRIUM IN REPEATED GAMES

### 2.1 Repeated Games with Partial Monitoring and Complete Information About the Stage Game

Let $N = \{1, \cdots, n\}$ be a set of players. Every player $i$ has a finite set $X_i$ of available actions. Let $X = X_1 \times, \cdots \times X_n$ be the set of action profiles, $\mathbf{x} = (x_1, \cdots, x_n)$. A game $\Gamma = \Gamma(\mathbf{u})$ is defined by a vector $\mathbf{u} = (u_1, \cdots, u_n)$ of payoff functions, where $u_i : X \to \Re$. In a repeated game associated with the 1-stage game $\Gamma(\mathbf{u})$, at each stage $t \geq 1$ every player $i$ chooses $x_i(t)$, and hence a *play*, which is an infinite sequence of action profiles $[x(1), x(2), \cdots]$ is generated, where for every $t$ $x(t) = (x_1(t), \cdots, x_n(t))$.

This play generates a stream of payoffs for each agent $i$, $[u_i(x(1)), u_i(x(2)), \cdots]$. Hence, in order to specify the total payoff of every player in the repeated game one must specify the way every player $i$ evaluates streams of payoffs. There are a few common ways to do it. In this paper we assume that for every stream of payoffs $r = [r_1, r_2, \cdots]$, the total payoff of $i$ is defined by the payoff function $V_i$ as follows:

$$V_i([r_1, r_2, \cdots]) = \liminf_{T \to \infty} \frac{1}{T} \sum_{t=1}^{T} r_t. \quad (1)$$

The second ingredient in defining a repeated game is to define the information available to every player $i$ at stage $t$ before she chooses $x_i(t)$. Let $X^t$, $t \geq 1$ be the set of all possible histories of action profiles of length $t$. Thus, $X^t = X \times X \cdots \times X$, $t$ times. A typical member of $X^t$ is denoted by $h^t$. Hence, every $h^t \in X^t$ has the form:

$$h^t = (x(1), x(2), \cdots, x(t)).$$

We call an element in $X_i^t = X_i \times X_i \times \cdots \times X_i$, $t$ times, an action history for player $i$.

At stage 1 every player $i$ just chooses $x_i(1) \in X_i$, and therefore the players jointly generate the first action profile $x(1) = (x_1(1), \cdots, x_n(1))$. After stage 1 is over, every player $i$ receives a signal, $s_i(1) \in S_i$, where $S_i$ is a given set of signals player $i$ can receive. The signal is produced through a pre-specified monitoring device $I_i^1 : X \to S_i$. Based on $s_i(1)$, and on his previous move $x_i(1)$ player $i$ chooses $x_i(2)$. Therefore the players jointly generate the profile of actions $x(2) = (x_1(2), \cdots, x_n(2))$. Before choosing their third actions, every player $i$ receives a new signal $s_i^2$, through her monitoring device $I_i^2$. Hence, $s_i(2) = I_i^2(x(1), x(2))$, and she chooses a new action, $x_i(3)$, based on her available information, $(s_i(1), s_i(2), x_i(1), x_i(2))$, and so on: At stage $t$, after the players generated the history of actions $h^{t-1} = (x(1), x(2), \cdots, x(t-1))$, player $i$ receives the signal $s_i(t-1) = I_i^{t-1}(h^{t-1})$. Player $i$ adds this recently received signal to the previous signals he received in previous stages and form a sequence of signals $(s_i(1), \cdots, s_i(t-1))$. Based on this sequence of signals, and on his past moves $(x_i(1), x_i(2), \cdots, x_i(t-1))$ $i$ chooses $x_i(t)$. Hence, the players jointly generate $x(t)$.[1]

**To summarize:** Fix the set of players $N$, and the strategy sets $X_i$. A repeated game $G = (\Gamma(\mathbf{u}), S_1, S_2, \cdots, S_n, I_1, I_2, \cdots, I_n)$ is defined by a 1-stage game $\Gamma(\mathbf{u})$, sets of signals $S_i$, and by monitoring devices $I_i$, where $I_i = (I_i^t)_{t=1}^{\infty}$ with $I_i^t : X^t \to S_i$.

For example: If for every player $i$ $I_i^t(h^t) = x(t)$ the associated repeated game has a *perfect monitoring of actions*.

---

[1] Note that our description implies that every $i$ has a *perfect recall*: At each stage he remembers his previous signals and his previous actions.

On the other extreme hand, if $I_i^t(h^t) = *$ constantly (that is, for every history of action profiles of any length, $i$ receives the same signal, $*$) there is no monitoring at all. We call such a monitoring device *trivial*.

In the next section we will deal with repeated first-price auctions. We will use the monitoring device, which after each stage announces the winning bid and the number of players who submitted the winning bid. This monitoring device, which will be identical to all players is less informative than the one announcing all bids, but of course it is more informative than the trivial device.

Similarly to histories of actions we denote by $S_i^t = S_i \times S_i \times \cdots \times S_i$, $t$ times the set of signal histories of length $t$ for player $i$, and by $S^t$ the set of joint signal histories of length $t$. The monitoring devices dictate the definition of strategies. A strategy of $i$ is a sequence of functions $f_i = (f_i^t)_{t=1}^{\infty}$, where $f_i^1 \in X_i$, and for every $t > 1$, $f_i^t$ is a function that assigns an action in $X_i$ for every pair of signals history $\delta_i^{t-1} \in S_i^{t-1}$ and history $h_i^{t-1} \in X_i^{t-1}$ for player $i$. Formally, $f_i^t : S_i^{t-1} \times X_i^{t-1} \to X_i$.

Let $\Sigma_i$ be the set of strategies of $i$, and let $\Sigma = \Sigma_1 \times \cdots \times \Sigma_n$ be the set of strategy profiles in the repeated game. As described above, every $\mathbf{f} = (f_1, \cdots, f_n) \in \Sigma$ defines recursively a play $x(f) = [x(f,1), x(f,2), \cdots] \in X^{\infty}$ where $X^{\infty}$ is the set of all infinite sequences of action profiles. The total payoff of $i$ in the repeated game when every player $j$ chooses $f_j$ is:

$$U_i(f) = V_i[u_i(x(f,1)), u_i(x(f,2)), \cdots,]$$

where $V_i$ is defined at (1).

A profile $f$ is in *equilibrium* in the repeated game if for every player $i$

$$U_i(f_i, f_{-i}) \geq U_i(g_i, f_{-i}),$$

for every $g_i \in \Sigma_i$.

## 2.2 Repeated Games With Partial Monitoring and Incomplete Information

As in the previous section, let $N = \{1, \cdots, n\}$ be a set of players, and for every player $i$ let $X_i$ be his set of available actions. The players are about to play a repeated game which will be chosen out of a class of repeated games $G_w$ indexed by a parameter $w \in \Omega$. Each such $w$ is called a *state*, and hence $\Omega$ is the set of states. The parameter $w$ defines the payoff functions and the monitoring devices of the players. Therefore, $G_w = (\Gamma(\mathbf{u}(w,.)), (S_i)_{i=1}^n, (I_i(w,.))_{i=1}^n).^2$ If all players know the state, then they will play a repeated game with partial monitoring with complete information about the stage game. However, the fun begins when the players do not know the chosen state. On the other hand every player gets some initial partial information about the state. Let $I_i^0 : \Omega \to S_i$ describe the device generating the initial information for $i$. That is, when $w$ is chosen, $i$ receives the signal $I_i^0(w)$.

In most models in economics it is assumed that $\Omega$ is a set of $n$-length sequences $w = (w_1, w_2, \cdots, w_n)$, and that the initial information of $i$ is $w_i$. We will make the same assumption when we will deal with repeated first-price auctions. In the auction model, every $w_i$ represents the value of the item to be sold to player $i$.

After receiving their initial information the players will play the repeated game $G_w$, without knowing $w$. The collection $G = ((G_\omega)_{w \in \Omega}, (I_i^0)_{i=1}^n)$ defines a *repeated game with partial monitoring and incomplete information*. If a probability distribution $\mu$ on $\Omega$ is given, the players are engaged in a Bayesian repeated game ( see e.g., Aumann & Maschler (1995)). If such a probability is not given we say that the players are engaged in a pre-Bayesian repeated game (see e.g., Ashlagi, Monderer, & Tennenholtz (2006) for a discussion of one stage pre-bayesian games).[3]

Note that in a repeated game with complete information, player $i$ does not condition his initial choice $x_i(1)$ on any information. In contrast, in $G$, $i$ bases his initial choice as well as any future choice on the initial information. Hence a strategy for $i$ in $G$ is a sequence $f_i = (f_i^t)_{t=1}^{\infty}$, such that $f_i^1 : S_i \to X_i$, and for every $t \geq 2$, $f_i^t : S_i^t \times X_i^{t-1} \to X_i$. The total payoff of $i$ in the repeated game with incomplete information $G$ when every player $j$ chooses the strategy $f_j$ is

$$U_i(\omega, f) = V_i[u_i(\omega, x(f,1)), u_i(\omega, x(f,2)), \cdots,]$$

where $V_i$ is defined at (1).

The main goal of the players in the repeated pre-Bayesian game is to maximize their payoffs. However, their payoffs depend on the true state, which they do not know, and on the strategies of the other players, which they do not know as well. If there is only one player, then a good optimizing strategy will be one that will give him his optimal total payoff at every state (whether he eventually knows the state or not). Brafman and Tennenholtz introduced the concept of learning equilib-rium into pre-Bayesian repeated games. Following their idea[4] we define a learning equilibrium as follows. Let $(f_1, f_2, \cdots, f_n)$ be a strategy profile in the repeated game with incomplete information describe above. For every $\omega \in \Omega$ we define by $f_i^w$ the induced strategy of $i$ in the repeated game $G_w$. That is, $f_i^w(s_i(1), s_i(2), \cdots, s_i(t-1), x_i(1), \cdots, x_i(t-1)) = f_i(I_i^0(\omega), s_i(1), s_i(2), \cdots, s_i(t-1), x_i(1), \cdots, x_i(t-1))$. A strategy profile in the pre-Bayesian game is a *learning equilibrium* if for every state $w$ the profile $(f_1^w, f_2^w, \cdots, f_n^w)$ is an equilibrium in the game defined by $w$, $G_w$.[5]

---

[2] There is no loss of generality in assuming the set of potential signals does not depend on the state. One can always define this set as the union of all set of signals in the system.

[3] Note that the term pre-Bayesian does not mean that we are about to assign probabilities to the states. It is just a convenient way to describe the situation to experts in Bayesian games: A pre-Bayesian game is a Bayesian game without the specification of a prior probability.

## 2.3 Robust Learning Equilibrium

Let $G$ be a repeated game with incomplete information as described above. We are about to define a learning equilibrium which is immune to strategic "mistakes". Let $f_i$ and $g_i$ be a couple of strategy profiles for player $i$ and let $T$ be a positive integer. We denote by $<g_i, f_i>_T$ the strategy for player $i$ in which she uses the strategy $g_i$ in the first $T$ rounds and the strategy $f_i$ in the rest of the game. That is, $<g_i, f_i\rangle>_T = <(g_i^t)_{t=1}^T, (f_i^t)_{t=T+1}^\infty>$.

We say that a strategy profile $f = (f_1, f_2, \ldots, f_n)$ is a *robust learning equilibrium* in $G$ if for every $T$ and for every strategy profile $g = (g_1, g_2, \ldots, g_n)$ the strategy profile $(<g_1, f_1>_T, <g_2, f_2>_T, \ldots, <g_n, f_n>_T)$ is a learning equilibrium in $G$.

The following example demonstrates the notion of a robust learning equilibrium.

### Example 2.1

|   | a   | b   | n   |   |   | a   | b   | n   |
|---|-----|-----|-----|---|---|-----|-----|-----|
| a | 1,1 | 6,0 | 0,0 |   | a | 5,5 | 0,6 | 0,0 |
| b | 0,6 | 5,5 | 1,0 |   | b | 6,0 | 1,1 | 0,0 |
| n | 0,0 | 0,0 | 0,0 |   | n | 0,0 | 0,0 | 0,0 |
|   |     | A   |     |   |   |     | B   |     |

*The set of states is $\Omega = \{A, B\}$. The row and column players are indexed by 1 and 2 respectively. The signals at the beginning of the game are: $I_1^0(A) = I_1^0(B)$, $I_2^0(A) = A$ and $I_2^0(B) = B$. Thus the column player knows the game which is played, while the row player does not know which game is played. The monitoring devices reveal to the players their own payoffs after each round. That is, for every history $h^t \in X^t$, $I_i^t(\omega, h^t) = u_i(\omega, x(t))$ for $i = 1, 2$ and for every $\omega \in \Omega$.*

*Consider the following strategy $f$ for player 1: Begin by playing a, regardless of the initial signal. If the history of payoffs is $1, 5, 1, 5...$ then play b if the last element in the history is 1, and play a if the last element in the history is 5. If the history of payoffs is $5, 1, 5, 1, ...$ then play a if the last element in the history is 1, and play b if the last element in the history is 5. In all other cases play n. Observe that $f$ is also a strategy for player 2. We claim that the profile $(f, f)$ is a learning equilibrium. If both players follow this profile then the total payoff for each of them is 3 at each of the two games. If player 1 deviates then in game A his total payoff will be at most 1 and in game B his total payoff will be 0. Therefore, a deviation is not profitable for agent 1. The same arguments apply to possible deviations by player 2. However, we show that $(f, f)$ is not a robust learning equilibrium. Let $g$ be a strategy where a player chooses n at the first stage. We will show that $(<g, f>_1, <g, f>_1)$ is not a learning equilibrium. If both players stick to their strategies in $(<g, f>_1, <g, f>_1)$ then the total payoff of player 1 equals 0 at both of the games. However, if player 1 deviates to the strategy "play b forever" then at game A the total payoff of player 1 is 1.*

## 2.4 System Failures

In the previous section we defined robust learning equilibrium in repeated games with incomplete information. Robust learning equilibrium is immune to deviations from the agents' prescribed actions, but it is not necessarily immune against failures of the monitoring devices. In this section we address this issue.

In order to capture the notion of failures of the monitoring devices, i.e. receiving incorrect signals from the system, we use the following idea which is fully formalized below. Every $\omega \in \Omega$ determines a particular profile of monitoring devices (one for each player), which in turn determines the correct information to be communicated to each player. In order to deal with failures we replace each state $\omega$ with a set of states of the form $(\omega, J)$ where each $J$ is a particular (potentially) faulty monitoring device profile. That is, $J = (J_1, J_2, \ldots, J_n)$, where $J_i = (J_i^t)_{t=1}^\infty$. With each new state $(\omega, J)$ we associate a new game $G_{(\omega, J)}$ which is obtained from $G_\omega$ by replacing the original profile of monitoring devices by $J$. This process defines a new and larger repeated game with incomplete information where we can test whether a given strategy profile is immune against the specified system failures.

Let $G = ((G_\omega)_{\omega \in \Omega}, (I_i^0)_{i=1}^n)$ be a repeated game with partial monitoring and incomplete information where for every $\omega \in \Omega$, $G_w = (\Gamma(\mathbf{u}(w, .), (S_i)_{i=1}^n), (I_i(w, .))_{i=1}^n)$.

Denote by $I(\omega)$ the profile of monitor devices $(I_1(\omega, \cdot), I_2(\omega, \cdot), ..., I_n(\omega, \cdot))$, where $I_i(\omega, \cdot) =$

---

[4] The original definition of Brafman & Tennenholtz (2004) deals with learning equilibrium for particular types of monitoring devices, and they did not consider initial information.

[5] Technically, a learning equilibrium is an ex post equilibrium in the associated pre-Bayesian repeated game.

$(I_i^t(\omega, \cdot))_{t=1}^\infty$.

For every $\omega \in \Omega$ let $F_\omega$ be a set of profiles of monitoring devices that satisfies the following:

- $I(\omega) \in F_\omega$.

- For every $J \in F_\omega$ there exists an integer $T$ such that $J^t = I^t(\omega)$ for every $t \geq T$, where $J^t = (J_1^t, J_2^t, \ldots, J_n^t)$ and $I^t(\omega) = (I_1^t(\omega, \cdot), I_2^t(\omega, \cdot), \ldots, I_n^t(\omega, \cdot))$.

The second condition above implies that every $J(\omega) \in F_\omega$ behaves identically to $I(\omega)$ (the non-faulty profile of monitoring devices in $\omega$) after a finite number of rounds.

Let $F = (F_\omega)_{\omega \in \Omega}$. $F$ determines the possible failure patterns at each of the states in $\Omega$. We are about to define a new repeated game with incomplete information derived from $F$ and $G = ((G_\omega)_{\omega \in \Omega}, (I_i^0)_{i=1}^n)$, which we will denote by $G(F)$. We first define a new state space $\Omega'$; $\Omega' = \{(\omega, J) | \omega \in \Omega, J \in F_\omega\}$. Secondly, for every $(\omega, J) \in \Omega'$ we define the associated repeated game with complete information $G_{(\omega, J)} = (\Gamma(\mathbf{u}(w, \cdot)), (S_i)_{i=1}^n, J)$. We now define $G(F) = ((G_{\omega'})_{\omega' \in \Omega'}, (I_i^0)_{i=1}^n)$. Notice that in $G$ and $G(F)$ the players receive the same initial information. Observe that a strategy in $G$ is also a strategy in $G(F)$ and vice versa.

Let $G$ be a repeated game with partial monitoring and incomplete information with a set of sates $\Omega$. Let $F$ determine the possible failure patterns at each of the states in $\Omega$. We say that a profile of strategies $f$ in $G$ is an *F-robust learning equilibrium* if $f$ is a robust learning equilibrium in the game $G(F)$.

## 3 ROBUST LEARNING EQUILIBRIUM IN AUCTIONS

We consider $n$ potential buyers, $N = \{1, 2, \cdots, n\}$, a seller, and a particular good. At every stage $t \geq 1$ the seller is running a first-price auction in order to sell a single unit of that good. Every buyer $i$ has a fixed valuation, $v_i \in V = \{1, 2, \cdots, m\}$, for that good. That is, $v_i$ is $i$'s maximal willingness to pay for the good. We define the set of states $\Omega$ to be the set of all vectors $v = (v_1, \ldots, v_n) \in V^n$. Each state $v$ determines a one-stage game, $\Gamma_v$ as follows. The action set of each player $i$ is $X_i = V$. This is the set of bids that are allowed in the auction. In order to define the payoff functions we use the following notations. For every profile of bids $x \in X = X_1 \times X_2 \cdots \times X_n$ let $M(x)$ be the maximal bid in $x$, i.e. $M(x) = \max_{i \in N} x_i$ and let $N(x)$ be the set of all players $i$ with $x_i = M(x)$. The payoff function of $i$, $u_i(v, x_1, x_2, \cdots, x_n)$, is defined as follows:

$$u_i(v, x) = \begin{cases} \frac{v_i - M(x)}{|N(x)|} & x_i = M(x) \\ 0 & otherwise. \end{cases}$$

Note that this definition is a convenient way to model a first-price auction with fair randomization as the tie breaking rule. The payoff of $i$ defined above, represents his expected payoff with respect to this tie breaking rule.

By Hon-Snir, Monderer, & Sela (1998) the following profile of bids is in equilibrium in $\Gamma_v$: $x_i = 1$ for every $i$ such that $v_i = 1$, and for every $i$ with $v_i > 1$,

$$x_i = \begin{cases} v_i - 1 & i \notin N(v) \\ v_i - 1 & i \in N(v) \text{ and } |N(v)| > 1 \\ v_{(2)} & i \in N(v) \text{ and } |N(v)| = 1, \end{cases} \quad (2)$$

where $v_{(j)}$ is the $j^{th}$ largest valuation among $v_1, \ldots, v_n$.

In order to define the corresponding repeated game $G_v$, we have to define the monitoring devices. In our setting all players have the same monitoring device, and the monitoring devices are state independent. Let $S_i = V \times \{1, 2, \ldots, n\}$. For every history of bid profiles $h_t = (x(1), x(2), \ldots, x(t)) \in X^t$ of length $t$, $I_i^t(h_t) = (M(x(t)), |N(x(t))|)$. That is, every player $i$ sees after round $t$ the wining bid and the number of players that made this bid.

To complete the definition of the repeated game with incomplete information, which we denote by $G$ we need to describe the structure of the initial information: Given the state $v = (v_1, v_2, \ldots, v_n)$ the initial information received by player $i$ is $v_i$.

Several algorithms (strategies) for playing in $G$ were presented in Hon-Snir, Monderer, & Sela (1998). It was shown that when all players adopt these algorithms the sequence of profiles of bids converges to the equilibrium defined in (2), of the one stage game $\Gamma_v$, where $v$ is the true state. The sequence converges in the following strong sense: There exist a time $T$ such the profiles of bids from time $T$ on is the equilibrium defined by (2). However, the issue of whether the related algorithms form a learning equilibrium was not discussed.

Let $\phi = (\phi(t))_{t=2}^\infty$ be a non-decreasing sequence of positive integers such that: $1 \leq \phi(t) \leq t - 1$ for every $t \geq 2$. For every such $\phi$ we define a learning algorithm $f_\phi$. Roughly speaking, when using $f_\phi$, at every stage $t$ the player uses only the history observed in the last $\phi(t)$ stages.

We will first prove the following three theorems, and then deal with the issue of failures.

**Theorem 3.1** *For every $\phi$, if the players adopt $(f_\phi, f_\phi, \ldots, f_\phi)$ then after a finite number of stages, at each state $v$ the players will play the equilibrium of $\Gamma_v$ described in (2).*

**Theorem 3.2** *If $\phi(t) \to \infty$ when $t \to \infty$ then $(f_\phi, f_\phi, \ldots, f_\phi)$ is a learning equilibrium.*

**Theorem 3.3** *If $\phi(t) \to \infty$ when $t \to \infty$, and also $(t - \phi(t))_{t=2}^\infty$ is a non-decreasing sequence converging to $\infty$, then $(f_\phi, f_\phi, \ldots, f_\phi)$ is a robust learning equilibrium.*

We now define the MaxBid algorithm $f_\phi$. We define the algorithm for an arbitrary player $i$.

**The MaxBid Algorithm - $f_\phi$:**

Recall that at stage $t$ player $i$'s signaling history is a sequence

$$\delta_i^{t-1} = ((M_k, |N_k|), (M_{k+1}, |N_{k+1}|), \ldots (M_{t-1}, |N_{t-1}|)),$$

where $k = t - \phi(t)$, $M_j$ denotes the maximal bid in stage $j$, and $|N_j|$ denotes the number of players that made that bid in stage $j$. At stage $t$ player $i$ will base his bid on $\delta_i^{t-1}$ and on the history of his actions

$$h_i^{t-1} = (x_i(k), \ldots, x_i(t-1)).$$

Let $\mathcal{M}_i(t) = \mathcal{M}_i(t, \delta_i^{t-1}, h_i^{t-1}) = \max_{t-\phi(t) \leq j \leq t}\{M_j | x_i(j) < M_j \text{ or } |N_j| > 1\}$. That is, $\mathcal{M}_i(t)$ is the observed maximal bid of all other players in the last $\phi(t)$ stages. However, if player $i$, and only player $i$, submitted the highest bid in all last $\phi(t)$ stages then $\mathcal{M}_i(t)$ is defined to be 1.

- At stage $t = 1$ choose $x_i(1) = 1$.
- At stage $t, t \geq 2$: for every signaling history $\delta_i^{t-1}$ and for every history of the player's actions $h_i^{t-1}$:

  If $v_i = 1$ let $x_i(t) = 1$.

  For $v_i > 1$ and $\mathcal{M}_i(t) = 1$ let $x_i(t) = \min[\max(x_i(t-1) - 1, 1), v_i - 1]$.

  For $v_i > 1$ and $\mathcal{M}_i(t) > 1$ let $x_i(t) = \min(\mathcal{M}_i(t) + 1, v_i - 1)$.

An immediate observation from the MaxBid algorithm is:

**Observation 3.4** *In the MaxBid alogrithm:(i) If $v_i = 1$ then $x_i(t) = 1$ for every $t \geq 1$. (ii) If $v_i \geq 2$ then $x_i(t) \leq v_i - 1$ for every $t \geq 1$.*

In the following example we illustrate the MaxBid algorithm in a 2-player repeated first price auction with incomplete information.

**Example 3.5** *Two players, both using MaxBid with an arbitrary $\phi$:*

The players' types: $\quad v_1 = 7 \quad v_2 = 5$

Player 1's bids: $\quad 1,2,3,4,5,5,\ldots$

Player 2's bids: $\quad 1,2,3,4,4,4,\ldots$

*The total payoff for the first player is 2 and for the second player is 0.*

Observe that in Example (3.5) the players indeed reach after five rounds the equilibrium of the one-stage auction given in (2).

**Proof of Theorem 3.1:** Let $v = (v_1, v_2, \ldots, v_n)$. We assume that there is at least one player with a valuation larger than 1 (otherwise, the proof is trivial). If $v_i = 1$ then player $i$ bids 1 forever, as required by (2). Let $v_i \geq 2$ where $v_i$ is not the maximal type in $v$. Player $i$ will raise her bid until she reaches $v_i - 1$, and then will continue submitting $v_i - 1$ indefinitely. If $v_i \geq 2$ and $v_i$ is the maximal type in $v$, and $|N(v)| = 1$ then $i$ will increase her bid (by 1) every round until her bid reaches $v_{(2)}$; if $|N(v)| > 1$ then player $i$ will raise her bid until $v_i - 1$, as required by (2). $\square$

We now show that not for every $\phi$, $(f_\phi, f_\phi, \ldots, f_\phi)$ is a learning equilibrium. In Example 3.6 we show that if $\phi(t) = 1$ for every $t \geq 2$ then the profile in which all players use the MaxBid algorithm is not a learning equilibrium.

**Example 3.6** *Consider the case $n = 2$ and let $\phi(t) = 1$ for every $t \geq 2$. In order to prove that $(f_\phi, f_\phi)$ is not a learning equilibrium, we will show that player 2 has a deviating strategy which makes him better off in the state $v = (v_1, v_2) = (7, 5)$. In the deviating strategy player 2 submits the following sequence of bids regardless of his type and the information available to him: $1,1,3,1,1,3,1,1,3,\ldots$. If both players adopted the $f_\phi$ then player 2's payoff would be 0 from a certain stage on, and therefore his total payoff in the game is 0. When player 2 uses the deviating strategy, the following bid history will be generated:*

Player 1's bids: $\quad 1,2,2,4,3,2,4,3,2,4,3,\ldots$

Player 2's bids: $\quad 1,1,3,1,1,3,1,1,3,1,1,\ldots$

*In this situation every $3^{rd}$ round the second player wins the one-stage auction and her total payoff in the repeated game is $\frac{2}{3}$.*

It is possible to construct examples in the spirit of Example 3.6 such that for any bounded sequence $\phi = (\phi(t))_{t=2}^\infty$, $(f_\phi, f_\phi, \ldots, f_\phi)$ is not a learning equilibrium.

We now prove Theorem 3.2, in which we provide a sufficient condition on $\phi$ such that $(f_\phi, f_\phi, \ldots, f_\phi)$ will be a learning equilibrium.

**Proof of Theorem 3.2:** Let $\phi(t) \to \infty$. Recall that $\phi(t)$ is non-decreasing in $t$. Let $f_j = f_\phi$ for every $j \in N$. Let $i \in N$. We need to show that $U_i(v, g_i, f_{-i}) \leq U_i(v, f)$ for every strategy $g_i$ and for every $v \in V^n$. The following claim will be useful:

**Claim 1:** Let $t \geq 2$. $\sum_{k=t}^{t+\phi(t)} u_i(v, x((g_i, f_{-i}), k)) \leq m^2 + U_i(v, f)\phi(t)$ for every strategy $g_i$ and for every $v \in V^n$.

*Proof of claim 1 (sketch):* Let $t \geq 1$, and let $x(t), x(t+1), \ldots x(t+\phi(t))$ be the sequence of action profiles generated by the strategy profile $(g_i, f_{-i})$. The proof follows from the following observations which hold for every $j \neq i$ and every integer $k, t \leq k < t + \phi(t)$:

1. If $x_j(k) < M(x(k))$ or $|N(x(k))| > 1$, and $x_j(k) < v_j - 1$, then $x_j(k+1) > x_j(k)$.

2. If $x_j(k) < x_j(k+1)$ then $x_j(q) \geq x_j(k+1)$ for every $k+1 \leq q \leq t + \phi(t)$.

If $v_i < v_j$ for some $j \neq i$, then $U_i(v, f) = 0$. Therefore, it is enough to show that $\sum_{k=t}^{t+\phi(t)} u_i(v, x((g_i, f_{-i}), k)) \leq m^2$. In order to show this it suffices to show that a player $i$, who uses $g_i$, can not a have positive payoff at more than $m$ stages in the interval between stage $t$ and stage $t + \phi(t)$. By the above observations, if $t_1 < t_2$ are two stages in which player $i$ has a positive payoff, then $x_j(t_2) > x_j(t_1)$, which implies that $i$ can not win at more than $m$ stages. The proof of the case where $i$'s type is maximal, is similar. □

We define the following sequence. Let $T_1 = 1$, and for every $k \geq 2$ let $T_k = T_{k-1} + 1 + \phi(T_{k-1} + 1)$.

The following holds:

$$U_i(v_i, g_i, f_{-i}) = \liminf_{T \to \infty} \frac{1}{T} \sum_{t=1}^{T} u_i(v, x((g_i, f_{-i}), t)) \leq$$

$$\liminf_{s \to \infty} \frac{1}{T_s} \sum_{k=2}^{s} \sum_{t=T_{k-1}}^{T_k} u_i(v, x((g_i, f_{-i}), t)) \leq$$

$$\liminf_{s \to \infty} \frac{sm^2 + U_i(v, f)T_s}{T_s}$$

where the first inequality holds since lim inf of a sequence is bounded above by any lim inf of any of its sub-sequences, and the second inequality follows from claim 1.

It suffices to show that $\lim_{s \to \infty} \frac{T_s}{s} = \infty$. Notice that $T_s = s + \Sigma_{k=2}^{s} \phi(T_{k-1} + 1)$ for $s \geq 2$. Therefore, it suffices to show that $\frac{\Sigma_{k=2}^{s} \phi(T_{k-1}+1)}{s} \to \infty$ when $s \to \infty$.

Since $T_{k-1} \geq k - 1$ then $\phi(T_{k-1} + 1) \geq \phi(k) \to \infty$ when $k \to \infty$. Therefore $\lim_{k \to \infty} \phi(T_{k-1} + 1) = \infty$. It is well known that if the limit of a sequence is $\infty$ then the corresponding sequence of means also converges to $\infty$. Our result follows. □

Before proving Theorem 3.3 we show that even when $\phi(t) \to \infty$, $(f_\phi, f_\phi, \ldots, f_\phi)$ is not necessarily a robust learning equilibrium, if $t - \phi(t)$ does not go to $\infty$.

**Example 3.7** *Consider the case $n = 2$. Let $\phi(t) = t - 1$ for every $t \geq 2$. We know that $(f_\phi, f_\phi)$ is a learning equilibrium. In order to show that $(f_\phi, f_\phi)$ is not a robust learning equilibrium we need to provide a couple of strategies $g_1$ and $g_2$ for players 1 and 2 respectively such that the strategy profile $(< g_1, f_\phi >_T, < g_2, f_\phi >_T)$ is not a learning equilibrium for some $T$. Let $g_1$ be the strategy - "play 2 forever" and $g_2$ be the strategy "play 5 forever". Suppose the state is $v = (v_1, v_2) = (7, 3)$, and let $T = 1$.*

*The following bid history will be generated by $(< g_1, f_\phi >_2, < g_2, f_\phi >_2)$:*

*Player 1's bids:    2,6,6,6,6,...*

*Player 2's bids:    5,2,2,2,2,...*

*The above history of bids generates a total payoff of 1 to player 1; however, she could have bid 5 on every round and receive a total payoff of 2.*

**Proof of Theorem 3.3:**

Let $(g_1, g_2, \ldots, g_n)$ be a strategy profile and let $T_1 \geq 1$. We have to show that $(< g_1, f_\phi >_{T_1}, < g_2, f_\phi >_{T_1}, \ldots, < g_n, f_\phi >_{T_1})$ is a learning equilibrium.

Since $t - \phi(t)$ is non-decreasing and converging to $\infty$, we have that for every $t \geq 1$ there exists an integer $T(t)$ such that $t' - \phi(t') > t$ for every $t' \geq T(t)$. Therefore, for every $t$ the following holds: for every $t' \geq T(t)$, the bid of a player who uses $f_\phi$ at $t'$ does not depend on any information received before time $t$. In particular, this is true for $t = T_1$. Hence, the infinite sequence of bids generated by $(< g_1, f_\phi >_{T_1}, < g_2, f_\phi >_{T_1}, \ldots, < g_n, f_\phi >_{T_1})$ starting from $T(T_1)$ coincides with the bids generated by $(f_\phi, f_\phi, \ldots, f_\phi)$. Therefore, the proof of Theorem 3.2 implies the desired result. □

**Corollary 3.8** *Let $\phi$ be as in Theorem 3.3, let $g$ be a strategy profile and let $T \geq 1$ be some integer. If every player $i$ adopts the strategy $< g_i, f_\phi >_T$ then after a finite number of stages, at each state $v$ the players will play the equilibrium of $\Gamma_v$ as in (2).*

The proof follows from Theorem 3.1, and the proof of Theorem 3.3.

**System failures in auctions:**

Let $G = ((G_\omega)_{\omega \in \Omega}, (I_i^0)_{i=1}^n)$ be a game with incomplete information, where for every $\omega \in \Omega$, $G_w = (\Gamma(\mathbf{u}(w,.)), (S_i)_{i=1}^n, (I_i(w,.))_{i=1}^n)$. In such games the monitoring devices at $\omega$ depend on $\omega$. However, in our setting of first-price auction with incomplete information the monitoring devices are state independent. This state-independent profile of monitoring devices is denoted by $I$. Therefore, in order to incorporate system failures into the first-price auction setting, we consider a set of possible monitoring devices which is also state independent. That is, $F_{\omega_1} = F_{\omega_2}$ for every $\omega_1, \omega_2 \in \Omega$. In the first-price auction setting a generic state is denoted by $v$ (and not by $\omega$). For an arbitrary state $v$, let $F_v$ be the set of all profiles of monitoring devices $J$ such that there exists an integer $T$ for which $J^t = I^t$ for every $t \geq T$. That is, in the context of first-price auction we allow any possible failure pattern satisfying the standard constraint that the system will re-cover from failure after finite time.

**Theorem 3.9** *If $\phi(t) \to \infty$ when $t \to \infty$, and also $(t-\phi(t))_{t=2}^\infty$ is a non-decreasing sequence converging to $\infty$, then $(f_\phi, f_\phi, \ldots, f_\phi)$ is an F-robust learning equilibrium.*

The proof of the above theorem is similar to the proof of Theorem 3.3, and is omitted from this version. Also omitted is a generalization of Corollary 3.8 to the model, which allows system failures.

## 4 Efficiency and Applicability

The paper introduces an approach to learning in multi-agent systems and apply it to the context of auctions. We consider multi-agent settings, where an organizer provides algorithms, to be followed by rational agents. Therefore, in the context of incomplete information, where learning is needed, the requirement that the learning algorithms provided will be in equilibrium is appealing. Our model might be criticized for using infinite horizon average payoffs. One may wish that convergence to desired outcomes will be efficient, as well as require that a deviation will become non-beneficial after a short time; these issues have been raised in earlier work on learning equilibrium (Brafman & Tennenholtz (2004)). These efficiency properties do hold in the auction setting studied in this paper, which makes our results applicable also to "impatient" agents. In addition, in order to have applicable results, one has to deal also with "non-strategic" deviations. The reader should notice that we treat the issue of agents' non-strategic deviations by using sub-game perfection, following the spirit of other definitions in the game-theory literature. However, we also consider failures of the monitoring device, an issue which to the best of our knowledge was not discussed in previous work. Our assumptions here refer to failures of the environment, and the way these effect strategic behavior, assuming the environment will eventually stabilize.